\documentstyle[11pt,moriond,epsfig,amsmath]{article}

\def\Journal#1#2#3#4{{#1} {\bf #2}, #3 (#4)}

\def\NPB{{\em Nucl.\ Phys.}\ B}
\def\PLB{{\em Phys.\ Lett.}\  B}

\def\PRD{{\em Phys.\ Rev.}\ D}
\def\ZPC{{\em Z.\ Phys.}\ C}
\def\EPJC{{\em Eur.\ Phys.\ J.}\ C} 
\def\JHEP{{\em J.\ High\ Energy\ Phys.}}

\def\lqcd{\Lambda_{\mbox{\scriptsize QCD}}}
\def\half{\mbox{\small $\frac{1}{2}$}}

\def\cM{{\cal{M}}}
\def\cA{{\cal{A}}}

\def\cA{{\cal{A}}}

\def\cP{{\cal{P}}}

\def\as{\alpha_{\mbox{\scriptsize s}}}
\def\PT{\mbox{\scriptsize PT}}
\def\NP{\mbox{\scriptsize NP}}

\def\be{\beta}

\def\prnt{}
\def\EEC{{\rm EEC}}
\def\NP{{\mbox{\scriptsize NP}}}
\def\PT{{\mbox{\scriptsize PT}}}

\begin{document}
\begin{flushright}
  hep-ph/9805323 \\
  IFUM-623-FT \\
  May 1998
\end{flushright}
\vspace*{2cm} 

\title{The Milan factor for jet-shape observables\footnote{Talk
    presented at ``Rencontres de Moriond, QCD and hadronic
    interactions,'' Les Arcs, France, April 1998, and at ``DIS98, 6th
    International Workshop on Deep Inelastic Scattering,'' Brussels,
    Belgium, April 1998.}}

\author{ G.P. Salam}

\address{INFN -- Sezione di Milano,\\ 
  Via Celoria 16, 20133 Milano, Italy}
               
\maketitle\abstracts{ This talk discusses and explains
  the two-loop corrections, also known as the Milan
  factor, for power corrections to jet-shape
  observables.\\
  }

\section{Introduction}
This talk aims to give a qualitative discussion of the issues related
to the two-loop calculations of power-suppressed contributions to
event-shape variables.\cite{DLMSmilan} For simplicity, it will
concentrate mostly on the situation in $e^+e^-$. For the one- and
two-loop DIS results, the reader is referred to the work of Dasgupta
and Webber.\cite{DWA2}

The essential issue in the phenomenological study of power-suppressed
contributions to QCD observables is that of universality, namely that
the leading power correction to any given observable $V$ can be
expressed in terms of a (perturbatively) calculable coefficient $c_V$
multiplied by an unknown (non-perturbative) number $\cA_{2p,q}$. This
latter number can be thought of as a particular moment of the
non-perturbative part of the QCD coupling:
\begin{equation}
  \label{eq:A2pq}
  \cA_{2p,q} = 
  \frac12
  \int_0^\infty
  \frac{dk^2}{k^2}\,k^{2p} \, \ln^q\frac{k^2}{\mu^2}\,
  \as^{\NP}(k^2)\,.
\end{equation}
Observables can be divided into classes according to the relevant
values of $p$ and $q$. For example many event-shape variables have
$2p=1$, $q=0$, and so are expected to have a power-suppressed
contribution $\cP$,
$$
\cP = \frac{4C_F}{\pi^2} \frac{c_V \cA_{1,0}}{Q},
$$
where $Q$ is the hard scale of the process. So in some sense
universality is the statement that for a certain class of observables, 
the relative coefficients of the power corrections are all
calculable. Experimentally, at the $20\%$ level, there is already some
fair evidence for universality among event-shape
observables.\cite{expee,exptalks} 

The rest of this talk will discuss the calculability of the
coefficients $c_V$, in particular the problems associated with the
1-loop calculation of power corrections and their solution with 
a two-loop calculation.

\section{Power corrections at one loop}

\begin{figure}[htbp]
  \begin{center}
    \scalebox{0.6}{\input{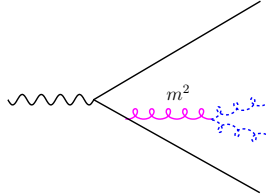}}
    \caption{Triggering on a gluon mass}
    \label{fig:gdec}
  \end{center}
\end{figure}

The methods for calculating power corrections at 1-loop accuracy have
reached a point of some maturity (and controversy). This talk will
discuss calculations in the particular context of the dispersive
approach,\cite{DMW} though a number of other approaches are
available.\cite{BB95,StKo,Zakh} In general one introduces a gluon mass
$m$, and looks for non-analyticity in $m^2$. In reality one should
understand this gluon mass as in figure~\ref{fig:gdec}, namely as the
virtuality of a gluon which then decays into two massless partons.
In a one-loop calculation, one has no information about the momenta of
these decay products, so the contribution to the event shape must be
taken as coming from the parent gluon. It was pointed out by Nason and
Seymour \cite{NS} that as a result of the partial non-inclusiveness of
event-shape variables this can be a rather poor approximation: the
decay products will not go in the same direction as the parent, and
therefore the parent and its decay products give different contributions
to the event shape.

Another manifestation of the same problem arises in the introduction
of the mass into the event-shape variable. Since the mass is fake, one 
is free to introduce it into the event-shape definition as one likes. If
one takes as an example the thrust variable,
$$
T = \max_{\vec{n}} \frac{\sum_i |\vec{p_i}.\vec{n}|}
{\sum_i |\vec{p_i}|}\,,
$$
one finds two suggestions for the power correction in the
literature. Beneke and Braun \cite{BB95} chose to include the effects
of the gluon mass in the denominator (i.e.\ truly the sum of the
modulus of the 3-momenta), whereas Dokshitzer and Webber \cite{DW}
chose not to.  Beneke and Braun obtained for the coefficient of the
power correction $c_T=-4G$, with $G$ being Catalan's constant, while
Dokshitzer and Webber obtained $c_T=-2$. At the one-loop level there is no
way to resolve this problem and one cannot unambiguously calculate the
coefficient $c_V$ of the power correction for a variable $V$:
universality is lost.

\section{The Milan factor}
\label{sec:mf}

The solution to the problem is to carry out a two-loop calculation. 
What one finds is that the two-loop result modifies the one-loop
result by a factor (known as the Milan factor \cite{WebbFras}). At
first sight this is a somewhat surprising result for a two-loop
correction. On the other hand, if the two-loop modification is to
resolve the one-loop ambiguity (a factor), then it has to be a factor
itself.  Another argument is that a
two-loop correction can always be recast as a change of scale:
\begin{equation}
  \label{eq:aa2}
  \as(Q^2) + c \as^2(Q^2) \simeq \as(A^2Q^2),  
\end{equation}
where  $A^2 = \exp(-4\pi c/\beta_0)$. But for a power correction, a
change of scale corresponds to a modification by a factor:
$$
\frac{\Lambda}{Q} \Rightarrow \frac{\Lambda}{AQ}\,.
$$
At this point one may start to wonder higher order
corrections will also be factors. The above argument suggests not,
since the addition of an $\as^3$
correction to \eqref{eq:aa2}, changes $A$ by
$$
A \to A(1+{\cal O}(\as/\pi))\,.
$$
The values obtained for the first-moment of the coupling suggest
that at small scales $\as/\pi \simeq 0.2$. This leads us to hope that
higher order corrections to the Milan factor will remain relatively
small ($\simeq 20\%$ ?), rather than a factor of order one.

One obvious worry if the two-loop corrections are factors is that
fits to data which worked reasonably well with old coefficients will
suddenly worsen. In fact most of the coefficients change by a common
factor $\cM\simeq 1.8$: given a suitable naive or one-loop definition
of the event-shape variable, typically in terms of the Sudakov
variables $\alpha$ and $\beta$, one finds that the entire variable
dependence of the power correction, both at one and two loops enters
through the quantity
$$
\rho_V = \int^{\infty}_{-\infty}
  d\eta \,f^{(V)}(\eta),
$$
with for example for thrust and the $C$-parameter
$$
f^{(T)}(\eta) = -e^{-|\eta|},
\quad f^{(C)}(\eta) = \frac{3}{\cosh \eta}\;.
$$
The rest of the calculation is variable independent. As a result the
relative coefficients of the power corrections stay the same at one
and two loops.

There are a few exceptions to this. The tables below summarise the old 
and new results. For $e^+e^-$ one has
\begin{center}
\begin{tabular}{|c||c||c||c|c||c|c|} \hline 
 $V$  & $ 1\!-\!T$ & $C$ & $M_T^2$ & $M^2_H$ & $B_T$ & $B_W$ \\ 
\hline
  old $c_V$      & 2 & --- & 2 & 2 & 2 & $2$ \\
\hline  
  new $c_V/\cM$  & 2 & $3\pi$ & 2 & 1 & 1 & $\half$ \\
\hline  
\end{tabular}
\end{center}
and for DIS,\cite{DWA2}
\begin{center}
\begin{tabular}{|c||c|c||c||c|} \hline 
 $F$       & $1-T_z$ & $1-T_c$ & $C_c$ & $\rho_Q$  \\ 
\hline
old $c_V$     & 2  &  2 &   $6\pi$  &   1     \\
\hline
new $c_V/\cM$ & 2  &  2 &   $3\pi$  &   1   \\
\hline  
\end{tabular}
\end{center}

\noindent In general the changes between the relative $c_V$ of different
variables can be attributed to the interplay between perturbative and
non-perturbative effects.\footnote{Except for $C_c$ in DIS where the
  change is due to the manner in which the gluon mass was included in
  the one-loop treatment. Note also that the coefficients given here
  for $1-T_z$ differ from those presented at the conference, which
  were wrong.}

For example in the original one-loop calculation to determine the
non-perturbative correction to the heavy-jet mass one considered only
a single gluon. The hemisphere which contained this gluon was
automatically the heaviest, and therefore the heavy-jet mass acquired
the whole non-perturbative contribution, and the light-jet mass none.
On the other hand, the formulation of the two-loop calculation takes
into account the presence also of perturbative radiation (though the
relevant techniques were first introduced for one-loop
calculations\cite{StKo,DWTD}). In general it is the perturbative
radiation which determines which jet is heavy, and since this is
uncorrelated with the non-perturbative radiation, the non-perturbative
``trigger'' gluon will be present in the heavy jet only half the time,
and therefore the heavy jet acquires half the correction of the total
jet mass.\footnote{As a result the coefficient for $M_H^2$ from the
  dispersive approach is now in agreement with that found earlier by
  Akhoury and Zakharov.\cite{AZ}}

A similar argument applies in the comparison of the total and wide-jet 
broadenings. Here though, there is the additional element of quark
recoil. The broadening is the sum of the transverse momenta (with
respect to the thrust axis) of all particles in the event:
$$
B_T = \frac{1}{2Q} \sum_i k_{T,i}
$$
In the presence of only a single non-perturbative gluon, the
transverse momentum of the recoiling quark is equal to that of the
gluon, and so the contribution to the broadening is twice that from the
gluon alone. But in practice perturbative radiation will give a
certain transverse momentum to the quark (somewhat larger than the
non-perturbative gluon transverse momentum), so that the effect of the
recoil from the non-perturbative gluon averages out to zero after the
integration over azimuth (or rather, gives a $(\ln Q)/Q^2$ type
correction). This halves one's naive expectation for the
non-perturbative contribution to the jet broadening. Analogous
considerations are expected to affect the predictions for the
jet-broadening in DIS.

It is to be recalled that the jet-broadenings are all expected to
exhibit a $(\ln Q)/Q$ type power correction, as opposed to $1/Q$ for
the other variables.

\section{Fractional power corrections}
\label{sec:fpc}
In this section we will examine the topic of fractional power
corrections, in particular for the energy-energy correlation (EEC),
defined as
$$
\EEC(\chi) = \sum_{ab} E_a E_b\; \delta(\cos\chi  -\cos\theta_{ab})\>.
$$
In the central region, namely $\chi \sim \pi/2$, the ratio of the
non-perturbative (NP) to perturbative (PT) contributions has the
behaviour
$$
\frac{\NP}{\PT} \sim \frac{1}{Q\sin \chi}\,.
$$
The configurations leading to this correction are those where one
triggers on a soft gluon and either the quark or anti-quark. The
divergence as $\sin \chi \to 0$ is closely connected to the
collinear divergence in gluon emission from a quark.

\begin{figure}[htbp]
  \begin{center}
    \epsfig{file=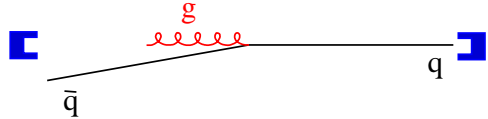,width=0.4\textwidth}
    \caption{Back-to-back EEC, showing the positions of the detectors
      and a typical configuration (only the non-perturbative gluon is
      shown)}
    \label{fig:eecpi}
  \end{center}
\end{figure}

The question of interest is what happens to the divergence when $\chi
\to \pi$. Figure \ref{fig:eecpi} illustrates a typical situation. The
choice $\chi=\pi$ means that the triggered gluon and quark are back to
back. Naively one would expect the gluon and the anti-quark to be
parallel, but in practice as a consequence of all-order perturbative
gluon emission, the quark and anti-quark are not quite
back-to-back,\cite{PPRW} so that there is a small angle $\bar{\chi}$
between the anti-quark and gluon, which when integrated over gives a
power correction (relative to the perturbative result)
$$
\frac{1}{Q} \int d^2{\bar\chi} \frac{P({\bar \chi})}{\sin {\bar \chi}}
\sim \frac{1}{Q} \cdot
\left(\frac{Q}{\lqcd}\right)^{\be_0/(\be_0+4C_F)} 
\propto  Q^{-0.372}\,.
$$
This estimate is the result of a steepest descent approach, which
appears to be too naive to produce the correct exponent, however
non-integer powers of $Q$ will definitely be there.\cite{EEC}
Analogous effects are expected for the energy-weighted particle sum in
the photon direction in the Breit frame in DIS, and for the height of
the plateau in the distribution of the transverse momentum of
Drell-Yan pairs.

\section{Merging}
\label{sec:merging}

So far a number of rather glib references have been made to
``non-perturbative gluons'' on the one hand and to ``perturbative
gluons'' on the other. This is connected with the use in \eqref{eq:A2pq}
of $\as^{\NP}$, the non-perturbative part of the coupling. In practice 
one defines $\as^{\NP}$ as the difference between the true coupling
and the perturbative expansion, $\as^{\PT}$, as used in the fixed-order
perturbative calculation, both in the CMW scheme:\cite{CMW}
$$
\as^{\NP}(k^2) \equiv \as(k^2) + \as^{\PT}(k^2)\,.
$$
One expects $\as^{\NP}$ to go to zero very rapidly, as the true and
the perturbative couplings coincide at moderate and large scales. As a 
result one chooses to truncate the moments \eqref{eq:A2pq} at an
infra-red matching scale $\mu_I$, so that for example $\cA_{1,0}$
becomes
\begin{eqnarray*}
  \label{eq:A10}
  \cA_{1,0} &\simeq& \int_0^{\mu_I} dk \, \as(k) - 
  \int_0^{\mu_I} dk \, \as^{\PT}(k)\\
   &\simeq& \mu_I\, \alpha_0(\mu_I) - \mu_I\left[
     \as(Q)
      - \beta_0\frac{\as^2}{2\pi}\left(\ln\frac{Q}{\mu_I}
        +1\right)
   \right]\,.
\end{eqnarray*}
On the second line, the second term, known as the merging part, has
been shown for the case of the two-loop perturbative
calculation.\footnote{Note that the merging term has been given for
  the case of a renormalisation scale $Q$; should one choose to use a
  renormalisation scale $\mu$ one should replace $Q$ with $\mu$. Note
  also that it has been given in the CMW scheme.} It is instructive to
examine its form for higher orders of perturbation theory. In general,
at $n^{\mbox{\scriptsize th}}$ order the merging piece will have the
form $\as^n n!$. This factorial divergence should cancel the
corresponding factorial (or renormalon) divergence in the fixed order
perturbative calculation which was the origin of the power correction
in the first place, leaving in the end a renormalon-free answer.

\section{Conclusions}
\label{sec:conc}
The main point of this talk has been that to determine the
coefficients for event-shape power corrections, and thus to be able to
test the concept of universality, it is not sufficient to perform a
one-loop determination of the power correction. This is because of the
non-inclusive nature of event shapes, which leads to an unresolvable
ambiguity in the result, according to one's choice of how to include
the fake dispersive gluon mass into the event-shape definition. 

The solution is to perform a two-loop calculation of the power
correction.\cite{DLMSmilan} As a result of the linearity of event-shape
variables (at least those with $1/Q$ leading power-suppressed
contributions) in the soft limit, it turns out that the variable
dependence enters both in the one- and two-loop calculations entirely
through one common simple integral; thus in one and two loops the
relative coefficients for event-shape power corrections stay the same
(as long as one chooses a suitable convention for the inclusion of the
gluon mass at one loop). 

For certain variables, the relative coefficient of the power
correction has changed: in the old 1-loop calculations there was
generally no simultaneous treatment of the perturbative and
non-perturbative gluon radiation, whereas the interaction between the
two is an essential part of the physics. One particularly interesting
example of the interaction between $\NP$ and $\PT$ physics is the
back-to-back energy-energy correlation where it leads to a fractional
power correction.

Finally, a critical element of all these discussions is that the power 
correction is not simply of the form $1/Q^p$ but that there is an
essential ``merging'' piece which subtracts out the renormalon
divergence in the perturbative calculation, ensuring that the final
answer is well defined.

\section*{Acknowledgements}
The results shown here for have been obtained in
collaboration with Yu.L. Dokshitzer, A. Lucenti and G. Marchesini.
It is a pleasure also to acknowledge illuminating discussions with
V.M. Braun, M. Dasgupta, G.P. Korchemsky, H.-U. Martyn, G. Sterman,
B.R. Webber and D. Wicke.

\end{document}